\documentstyle[amssymb,aps]{revtex}

\begin{document}
\author{E. J. Ferrer$\thanks{{\protect\footnotesize On leave from Department of
Physics, SUNY-Fredonia, NY 14063, USA}}^{1}$ and V. de la Incera$^{*2}$}
\address{$^{1}${\small Institute for Space Studies of Catalonia, CSIC, Gran Capita
2-4, 08034 Barcelona, Spain.}\\
$^{2}${\small University of Barcelona, Department of Structure and
Constituents of Matter,}\\
{\small Diagonal 647, 08028 Barcelona, Spain.}}
\title{Magnetic Catalysis in the Presence of Scalar Fields}
\maketitle

\begin{abstract}
The influence of scalar field interactions in the magnetic catalysis
phenomenom is studied in a gauge theory with scalar and fermion fields. It
is shown that the external magnetic field catalyzes the appearance of a
fermion-antifermion condensate along with a non-zero scalar vev, hence
generating a fermion dynamical mass and breaking the discrete chiral
symmetry of the theory. The model does not require the introduction of a
scalar mass term with a wrong sign in the original Lagrangian to generate
the scalar vev, nor does it exhibit dimensional transmutation ''a la
Coleman-Weinberg.'' The scalar vev noticeably enhances the dynamically
generated fermion mass. Possible cosmological applications of these results
are suggested.

{\small PACS: 11.30Qc, 11.30 Rd, 12.15-y. Keywords: non-perturbative
effects, external magnetic fields, symmetry breaking}
\end{abstract}

Symmetry behavior in quantum field theories under the influence of external
fields has long been a topic of intensive study in theoretical physics\cite
{books}. In the present paper we are particularly interested in
non-perturbative effects produced by external magnetic fields in gauge
theories with scalar-scalar and scalar-fermion interactions.

When a non-perturbative analytic approach is used to study field theories in
external magnetic fields, new non-trivial effects are found. An important
example of these non-perturbative effects is the formation of a chiral
symmetry breaking fermion condensate $<\overline{\psi }\psi >$ and of a
dynamically generated fermion mass in the presence of an external magnetic
field, known in the literature as magnetic catalysis\cite{mirans-gus-shoko},%
\cite{NJL-2dimen}. As was originally shown by Gusynin, Miransky and Shovkovy 
\cite{mirans-gus-shoko}, the magnetic field catalyzes the symmetry breaking
by reducing to the weakest attractive coupling the strength of the
interaction needed to break the symmetry. This phenomenon has proven to be
rather universal and model independent \cite{ackley}-\cite{composite}. The
magnetic catalysis is not only interesting from a purely fundamental point
of view, but it has potential application in condensed matter \cite{cond-mat}%
-\cite{vv3} and cosmology \cite{vv1}.

In the present paper we aim to investigate the influence of scalar field
interactions in the magnetic catalysis phenomenon. With this goal, we will
consider a simple model field theory containing self interacting scalar
fields, as well as Yukawa fermion-scalar interactions in the presence of an
external constant magnetic field. Using a self-consistent approach, we will
determine how the existence of these interactions in the presence of the
magnetic field can affect the stability of the vacuum. We will see below
that due to the magnetic field, a non-zero vacuum expectation value (vev) of
the scalar field and a fermion dynamical mass arise as the solution of the
minimum equations of the system, breaking in this way the discrete chiral
symmetry of the original Lagrangian. However, in contrast to more
conventional mechanisms to generate scalar vev's, the present model does not
require the introduction of a scalar mass term with a wrong sign in the
original Lagrangian, nor does it have dimensional transmutation ''a la
Coleman-Weinberg.'' Instead, the scalar vev is catalyzed, along with a
fermion-antifermion condensate, by the external magnetic field. The only
assumption we make is that all couplings are weak enough to justify the
Hartree-Fock approach used below.

Both the fermionic condensate and the scalar vev contribute to the
dynamically generated fermion mass. We find that the scalar vev and the
fermion dynamical mass grow with the square root of the magnetic field
strength. The main effect of the scalar interactions is a significant
enhancement of the magnetic catalysis as reflected in the increase of the
fermion dynamical mass compared to the case without scalars. The mass
enhancement resembles a similar one produced by lowering the number of
spatial dimensions in the Nambu-Jona-Lasinio (NJL) model from 3+1 to 2+1 
\cite{NJL-2dimen}. Like other theories with magnetic catalysis, no critical
magnetic field is required in the present model for the zero-temperature
chiral symmetry breaking to take place.

The studied model is not intended as a realistic theory, but rather as an
example of a large class of theories with scalar fields, on which a
dynamical symmetry breaking can be catalyzed by an external magnetic field.
In this sense, it could be useful for condensed matter, as well as for
cosmological applications.

Our starting point will be a theory of gauge, fermion and real scalar fields
described by the following Lagrangian density

\begin{equation}
L=-\frac{1}{4}F^{\mu \nu }F_{\mu \nu }+i\overline{\psi }\gamma ^{\mu
}\partial _{\mu }\psi +g\overline{\psi }\gamma ^{\mu }\psi A_{\mu }-\frac{1}{%
2}\partial _{\mu }\varphi \partial ^{\mu }\varphi -\frac{\lambda }{4!}%
\varphi ^{4}-\frac{\mu ^{2}}{2}\varphi ^{2}-\lambda _{y}\varphi \overline{%
\psi }\psi  \label{e1}
\end{equation}
It has a U(1) gauge symmetry, 
\begin{eqnarray}
A_{\mu } &\rightarrow &A_{\mu }+\frac{1}{g}\partial _{\mu }\alpha (x) 
\nonumber \\
\psi &\rightarrow &e^{i\alpha (x)}\psi ,  \label{ee1}
\end{eqnarray}
a fermion number global symmetry 
\begin{equation}
\psi \rightarrow e^{i\theta }\psi ,  \label{ee2}
\end{equation}
and a discrete chiral symmetry 
\begin{equation}
\psi \rightarrow \gamma _{_{5}}\psi ,\qquad \overline{\psi }\rightarrow -%
\overline{\psi }\gamma _{_{5}},\qquad \varphi \rightarrow -\varphi
\label{e2}
\end{equation}

Note that a fermion mass term $m\overline{\psi }\psi $ is forbidden, since
it is invariant under (\ref{ee1}) and (\ref{ee2}), but not under the
discrete chiral symmetry (\ref{e2}).

To study the vacuum solutions that could arise in the theory (\ref{e1})
under the influence of an external constant magnetic field $B$ (without loss
of generality we will assume throughout the paper that $sgn(gB)>0$), we need
to solve the extremum equations of the effective action $\Gamma $ for
composite operators\cite{jackiw},\cite{miransbook}

\begin{eqnarray}
\frac{\delta \Gamma (\varphi _{c},\overline{G})}{\delta \overline{G}} &=&0,\;
\label{e3} \\
\frac{\delta \Gamma (\varphi _{c},\overline{G})}{\delta \varphi _{c}} &=&0
\label{e4}
\end{eqnarray}
where $\overline{G}(x,x)=\sigma (x)=\left\langle 0\mid \overline{\psi }%
(x)\psi (x)\mid 0\right\rangle $ is a composite fermion-antifermion field,
and $\varphi _{c}$ represents the vev of the scalar field. If the minimum
solutions of (\ref{e3}) and (\ref{e4}) are different from zero the discrete
chiral symmetry (\ref{e2}) will be broken and the fermions will acquire
mass. We shall see that due to the magnetic field this is indeed the case.

We choose the composite field $\overline{G}(x,x),$ ignoring the second
possible one, $\pi (x)=\left\langle 0\mid \overline{\psi }(x)i\gamma
_{5}\psi (x)\mid 0\right\rangle ,$ since the effective action can be a
function only of the chirally invariant combination $\rho ^{2}=\sigma
^{2}+\pi ^{2}$.

The loop expansion of the effective action $\Gamma $ for composite operators 
\cite{jackiw},\cite{miransbook} can be expressed as

\begin{equation}
\Gamma \left( \overline{G},\varphi _{c}\right) =S\left( \varphi _{c}\right)
-iTr\ln \overline{G}^{-1}+i\frac{1}{2}Tr\ln D^{-1}+i\frac{1}{2}Tr\ln \Delta
^{-1}-iTr\left[ G^{-1}\left( \varphi _{c}\right) \overline{G}\right] +\Gamma
_{2}\left( \overline{G},\varphi _{c}\right) +C  \label{e5}
\end{equation}
Here $C$ is a constant and $S\left( \varphi _{c}\right) $ is the classical
action evaluated in the scalar vev $\varphi _{c}.$ The bar on the fermion
propagator $\overline{G}\left( x,y\right) $ means that it is taken full,
while the non-bar notation indicates free propagators, as it is the case for
the gauge propagator $D_{\mu \nu }(x-y)=\int \frac{d^{4}q}{(2\pi )^{4}}\frac{%
e^{iq\cdot (x-x^{\prime })}}{q^{2}-i\epsilon }\left( g_{\mu \nu }-(1-\xi )%
\frac{q_{\mu }q_{\nu }}{q^{2}-i\epsilon }\right) ,$ and the scalar one $%
\Delta (x-y)=\int \frac{d^{4}q}{(2\pi )^{4}}\frac{e^{iq\cdot (x-x^{\prime })}%
}{q^{2}+M^{2}-i\epsilon }$, with $M^{2}=\frac{\lambda }{2}\varphi
_{c}^{2}+\mu ^{2}.$ A dependence on full boson propagators is not included,
since we do not expect that any of the boson fields will acquire expectation
values of their composite operators. In general $\Gamma _{2}\left( \overline{%
G},\varphi _{c}\right) $ represents the sum of two and higher loop
two-particle irreducible vacuum diagrams with respect to fermion lines. For
a weakly coupled theory one can use the Hartree-Fock approximation, which
corresponds to retaining only that contribution to $\Gamma _{2}$ which is
lowest-order in coupling constants (i.e. two-loop graphs only). In this case 
$\Gamma _{2}$ is

\begin{eqnarray}
\Gamma _{2}\left( \overline{G},\varphi _{c}\right) &=&\frac{g^{2}}{2}\int
d^{4}xd^{4}ytr\left[ \overline{G}\left( x,y\right) \gamma ^{\mu }\overline{G}%
\left( y,x\right) \gamma ^{\nu }D_{\mu \nu }(x,y)\right]  \nonumber \\
&&-\frac{g^{2}}{2}\int d^{4}xd^{4}ytr\left( \gamma ^{\mu }\overline{G}\left(
x,x\right) \right) D_{\mu \nu }(x-y)tr\left( \gamma ^{\nu }\overline{G}%
\left( y,y\right) \right)  \nonumber \\
&&+\frac{\lambda _{y}^{2}}{2}\int d^{4}xd^{4}ytr\left[ \overline{G}\left(
x,y\right) \overline{G}\left( y,x\right) \Delta (x,y)\right]  \nonumber \\
&&-\frac{\lambda _{y}^{2}}{2}\int d^{4}xd^{4}ytr\left( \overline{G}\left(
x,x\right) \right) \Delta (x-y)tr\left( \overline{G}\left( y,y\right) \right)
\label{e6}
\end{eqnarray}

The extremum equations $\left( \ref{e3}\right) $ and $\left( \ref{e4}\right) 
$ correspond, respectively, to the Schwinger-Dyson (SD) equation for the
fermion self-energy operator $\Sigma $ (gap equation)$,\ $and to the usual
minimum equation for the expectation value of the scalar field, which in the
presence of the magnetic field has to be determined in a self-consistent
way, that is, simultaneously with the gap equation.

Although we have introduced a bare scalar mass $\mu $ in (\ref{e1}), as we
are interested in the possibility of a dynamically generated scalar vev, we
shall take the limit $\mu \rightarrow 0$ at the end of our calculations.

The second to fifth terms in the effective action $\left( \ref{e5}\right) $
correspond to one-loop contribution. Their evaluation is quite
straightforward (the scalar self-interaction can be renormalized in the
usual way), with the exception of the fermion contributions, which contain
the background magnetic field. Then, let us calculate explicitly the
one-loop fermion contribution coming from the term 
\begin{equation}
\Gamma _{f}^{(1)}=-iTr\left[ G^{-1}\left( \varphi _{c}\right) \overline{G}%
\right]  \label{e7}
\end{equation}
in (\ref{e5}). The trace operator is along continuous and discrete variables.

The full fermion propagator $\overline{G}\left( x,y\right) $ in the presence
of a constant magnetic field $B$ appearing in Eq. (\ref{e7}) can be written
as \cite{ritus},\cite{ackley}

\begin{equation}
\overline{G}\left( x,y\right) =\sum\limits_{k}\int \frac{dp_{0}dp_{2}dp_{3}}{%
\left( 2\pi \right) ^{4}}E_{p}\left( x\right) \left( \frac{1}{\gamma .%
\overline{p}+\Sigma (p)}\right) \overline{E}_{p}\left( y\right)  \label{e8}
\end{equation}

Similarly, the free fermion inverse propagator in the presence of $B,$
denoted by $G^{-1}\left( \varphi _{c}\right) ,$ is given by

\begin{equation}
G^{-1}\left( x,y,\varphi _{c}\right) =\sum\limits_{k^{\prime }}\int \frac{%
dp_{0}^{\prime }dp_{2}^{\prime }dp_{3}^{\prime }}{\left( 2\pi \right) ^{4}}%
E_{p^{\prime }}\left( x\right) \left( \gamma .\overline{p}^{\prime
}+m_{0}\right) \overline{E}_{p^{\prime }}\left( y\right) ,  \label{e9}
\end{equation}
with $\overline{p}=(p_{0},0,-\sqrt{2gBk},p_{3}),$ $k$ the Landau level
index, and $m_{0}=\lambda _{y}\varphi _{c}$ the fermion mass appearing after
the shift $\varphi \rightarrow \varphi +\varphi _{c}$ in the scalar field.
The magnetic field $B$ has been taken along the third axis$.$

In the above equations we have introduced the Ritus' $E_{p}$ functions \cite
{ritus}. These orthonormal and complete set of function-matrices provides an
alternative method to the Schwinger's approach to problems of QFT on
electromagnetic backgrounds\footnote{%
For an application of Ritus' method to the QED Schwinger-Dyson equation in a
magnetic field see ref.\cite{ackley}.}. Details on the $E_{p}$ functions
properties and the notation used below can be found in ref. \cite{vv1}.

Using Eqs. (\ref{e8}) and (\ref{e9}), and the orthonormality of the $E_{p}$
functions, the integrals in the coordinate variables, as well as the sum and
integrals in one of the momentum variables can be done straightforwardly to
find that the one-loop fermion contribution (\ref{e7}) can be expressed as

\begin{equation}
\Gamma _{f}^{(1)}=-i\left( 2\pi \right) ^{4}\delta
^{(3)}(0)\sum\limits_{k}\int \frac{d^{3}p}{\left( 2\pi \right) ^{4}}%
Tr\left\{ \frac{\gamma .\overline{p}+m_{0}}{\gamma .\overline{p}+\Sigma (%
\overline{p})}\right\}  \label{e10}
\end{equation}
where the notation $\delta ^{(3)}(k)=\delta (k_{0})\delta (k_{2})\delta
(k_{3})$ is understood.

At this point we need to consider the structure of the mass operator $\Sigma
\ $introduced in ref. \cite{vv2}

\begin{equation}
\mathop{\textstyle \sum }%
(\overline{p})=Z_{_{\Vert }}(\overline{p})\gamma \cdot \overline{p}_{_{\Vert
}}+Z_{\bot }(\overline{p})\gamma \cdot \overline{p}_{_{\bot }}+m(\overline{p}%
)  \label{e11}
\end{equation}
where $\overline{p}_{_{\Vert }}=(p_{0,}0,0,p_{3})$, $\overline{p}_{_{\bot
}}=(0,0,-\sqrt{2gBk},0).$ The coefficients $Z_{_{\Vert }}(\overline{p}),$ $%
Z_{\bot }(\overline{p})$ and $m(\overline{p})$ are functions of $\overline{p}%
^{2}.$ Note the usual separation in the presence of a magnetic field between
parallel and transverse variables. Substituting with the structure (\ref{e11}%
) in Eq. (\ref{e10}), taking the trace, and performing the Wick rotation to
Euclidean coordinates, we obtain, after integrating in $p_{2}$

\begin{equation}
\Gamma _{f}^{(1)}=8\pi gB\delta ^{(4)}(0)\sum\limits_{k}\int dp_{4}dp_{3}%
\frac{(1+Z_{_{\Vert }})\overline{p}_{\Vert }^{2}+(1+Z_{\bot })\overline{p}%
_{_{\bot }}^{2}+m(\overline{p})m_{0}}{(1+Z_{_{\Vert }})^{2}\overline{p}%
_{\Vert }^{2}+(1+Z_{\bot })^{2}\overline{p}_{_{\bot }}^{2}+m^{2}(\overline{p}%
)}  \label{e12}
\end{equation}

We are interested in the contribution of $\Gamma _{f}^{(1)}$ to the minimum
equations (\ref{e3}) and (\ref{e4}). In the case of the gap equation, such a
contribution can be found differentiating directly Eq. (\ref{e7}) with
respect to $\overline{G}.$ For the minimum of the scalar field, we need the
derivative of (\ref{e12}) with respect to $\varphi _{c}.$ It takes the form

\begin{equation}
\frac{\partial \Gamma _{f}^{(1)}}{\partial \varphi _{c}}=8\pi \lambda
_{y}gB\delta ^{(4)}(0)\sum\limits_{k}\int dp_{4}dp_{3}\frac{m(\overline{p})}{%
\overline{p}_{\Vert }^{2}+(1+Z_{\bot })^{2}\overline{p}_{_{\bot }}^{2}+m^{2}(%
\overline{p})}  \label{e13}
\end{equation}
where the solution\footnote{%
The demostration that $Z_{_{_{\Vert }}}=0$ is a solution of the gap equation
in the present theory can be done along the same line of reasoning followed
in the Appendix of the first paper of ref.\cite{vv1}.} $Z_{_{_{\Vert }}}=0$
of the SD equation (\ref{e3}) was explicitly used.

For large magnetic field, the main contribution to the sum in $k$ comes from
the lower Landau level, i.e. $\overline{p}_{_{\bot }}^{2}=2gBk=0$. Within
this approximation Eq. (\ref{e13}) becomes

\begin{equation}
\frac{\partial \Gamma _{f}^{(1)}}{\partial \varphi _{c}}=8\pi \lambda
_{y}gB\delta ^{(4)}(0)\int dp_{4}dp_{3}\frac{m(\overline{p}_{\Vert })}{%
\overline{p}_{\Vert }^{2}+m^{2}(\overline{p}_{\Vert })}  \label{e14}
\end{equation}

In general, the dynamical mass $m(\overline{p}_{\Vert })$ depends on the
momentum. Nevertheless, it is reasonable to expect that, similarly to QED 
\cite{mirans-gus-shoko},\cite{composite}, $m(\overline{p}_{\Vert })$ behaves
as a constant in the infrared region, and diminishes with increasing $\left| 
\overline{p}_{_{_{_{\Vert }}}}\right| $. Then the integral (\ref{e14}) is
dominated by the contributions from the infrared region $\left| \overline{p}%
_{_{_{_{\Vert }}}}\right| <\sqrt{gB}.$ Therefore, we can approximate the
function $m(\overline{p}_{\Vert })$ by a constant solution $m(\overline{p}%
_{\Vert })\approx m(o)=m\ll \sqrt{gB}$ and cut the momentum integration at $%
\sqrt{gB}$ to obtain

\begin{equation}
\frac{\partial \Gamma _{f}^{(1)}}{\partial \varphi _{c}}=V^{(4)}\frac{%
\lambda _{y}}{2\pi ^{2}}gBm\ln \left( \frac{gB}{m^{2}}\right)
=-V^{(4)}\lambda _{y}<\overline{\psi }\psi >  \label{e15}
\end{equation}
where $V^{(4)}$ represents an infinite four dimensional volume, and $<%
\overline{\psi }\psi >=iTr\left\{ \overline{G}(x,x)\right\} =-\frac{gBm}{%
2\pi ^{2}}\ln \left( \frac{gB}{m^{2}}\right) $ denotes the
fermion-antifermion condensate \cite{ackley},\cite{smilga} induced by the
external magnetic field.

The two-loop contributions are a little more involved. As we have not enough
space in a letter to give all the detailed calculations, we will explicitly
show, for the sake of understanding, the evaluation of one term. The others
can be found in a similar way. The complete calculation will be published
elsewhere.

First, notice that the second and fourth term in Eq. (\ref{e6}) generate
tadpole diagrams in the SD equation (\ref{e3}). It is easy to realize that
the tadpole diagram with the gauge-fermion vertex vanishes. However, the
tadpole associated to the scalar-fermion vertex is not zero when $\varphi
_{c}\neq 0$ and, as shown below, it has a significant contribution to the
gap equation. Let us evaluate this tadpole contribution, which we denote by $%
\sum\nolimits^{T}.$ 
\begin{equation}
\mathop{\textstyle \sum }%
\nolimits^{T}(x,y)=i\frac{\delta \Gamma _{2}^{^{T}}}{\delta G}=-i\lambda
_{y}^{2}\delta ^{4}(x-y)\int d^{4}z\Delta (x-z)tr\left[ \overline{G(}%
z,z)\right]  \label{e16}
\end{equation}

We can transform Eq. (\ref{e16}) to momentum space with the help of the $%
E_{p}\left( x\right) $ functions to obtain

\[
\int d^{4}xd^{4}y\overline{E}_{p}\left( x\right) 
\mathop{\textstyle \sum }%
\nolimits^{T}(x,y)E_{p^{\prime }}\left( y\right) =(2\pi )^{4}\widehat{\delta 
}^{(4)}(p-p^{\prime })%
\mathop{\textstyle \sum }%
\nolimits^{T}(\overline{p}) 
\]
\begin{eqnarray}
&=&-i\lambda _{y}^{2}\int d^{4}xd^{4}z\overline{E}_{p}\left( x\right) \int 
\frac{d^{4}q}{\left( 2\pi \right) ^{4}}\frac{e^{iq\left( x-z\right) }}{%
q^{2}+M^{2}+i\in }  \nonumber \\
&&\times \sum\limits_{k"}\int \frac{d^{3}p^{"}}{\left( 2\pi \right) ^{4}}%
Tr\left\{ E_{p"}\left( z\right) \left( \frac{1}{\gamma .\overline{p}"+\sum (%
\overline{p}")}\right) \overline{E}_{p"}\left( z\right) \right\}
E_{p^{\prime }}\left( x\right)  \label{e17}
\end{eqnarray}

and then integrate in $x$ and $z$ to find

\[
(2\pi )^{4}\widehat{\delta }^{(4)}(p-p^{\prime })\sum\nolimits^{T}(\overline{%
p})=-i\lambda _{y}^{2}\int d^{4}q\sum\limits_{k"}\int d^{3}p^{"}\delta
^{\left( 3\right) }(q)\delta ^{\left( 3\right) }(p^{\prime }+q-p) 
\]
\[
\times e^{-\widehat{q}_{\bot }^{2}}\frac{e^{iq_{1}(p_{2}^{\prime
}+p_{2}-2p_{2}")/2gB}}{q^{2}+M^{2}+i\in }\sum\limits_{\sigma }\frac{%
e^{i(n-n^{\prime })\varphi }}{\sqrt{n(k,\sigma )!n^{\prime }(k^{\prime
},\sigma )!}}J_{nn^{\prime }}(\widehat{q}_{\bot })\Delta (\sigma ) 
\]
\begin{equation}
\times \sum\limits_{\sigma "}\left\{ Tr\left[ \Delta \left( \sigma "\right) 
\frac{1}{\gamma .\overline{p}"+\sum (\overline{p}")}\right] \frac{1}{%
n"(k",\sigma ")!}J_{n"n"}(\widehat{q}_{\bot })\right\} ,  \label{e18}
\end{equation}
where $\widehat{q}_{\bot }^{2}=\left( q_{1}^{2}+q_{2}^{2}\right) /gB$, $%
\varphi \equiv \arctan \left( \frac{q_{2}}{q_{1}}\right) $, and $n(k,\sigma
)\equiv k+\frac{\sigma }{2}-\frac{1}{2}$.

This result can be further simplified by approximating the $J-$functions by
their small-$\widehat{q}_{\bot }^{2}$ limit 
\begin{equation}
J_{nn"}(\widehat{q}_{\bot })\equiv \sum\limits_{m=0}^{\min (n,n")}\frac{n!n"!%
}{m!(n-m)!(n"-m)!}[iq_{_{\perp }}]^{n+n"-2m}\rightarrow \frac{\left[ \max
(n,n")\right] !}{\left| n-n"\right| !}\left[ i\widehat{q}_{\bot }\right]
^{\left| n-n"\right| }\rightarrow n!\delta _{nn"},  \label{e19}
\end{equation}
which is justified by the presence of the exponential factor $e^{-\widehat{q}%
_{\bot }^{2}}$ in the integrand of Eq.(\ref{e18}). Now the sums in the $%
\sigma ^{\prime }s$ and the trace can be easily done, along with all but two
momentum integrals, to find the following expression in Euclidean space, 
\begin{equation}
\mathop{\textstyle \sum }%
\nolimits^{T}=\frac{\lambda _{y}^{2}}{2\pi ^{3}}\frac{gB}{M^{2}}%
\sum\limits_{k"}\int dp_{4}^{"}dp_{3}^{"}\frac{m(\overline{p}")}{\left(
1+Z_{_{_{\Vert }}}\right) ^{2}\overline{p}_{_{_{_{\Vert }}}}"^{2}+\left(
1+Z_{\bot }\right) ^{2}\overline{p}_{_{\bot }}"^{2}+m^{2}(\overline{p}")}
\label{e20}
\end{equation}

Just like in the one-loop case, it can be seen that the main contribution to
Eq. (\ref{e20}) comes from the $k"=0$ term of the sum. Thus, the tadpole
contribution to the gap equation reduces to

\begin{equation}
\mathop{\textstyle \sum }%
\nolimits^{T}=\frac{\lambda _{y}^{2}}{2\pi ^{3}}\frac{gB}{M^{2}}\int
dp_{4}^{"}dp_{3}^{"}\frac{m(\overline{p}_{_{_{\Vert }}}")}{\overline{p}%
_{_{_{_{\Vert }}}}"^{2}+m^{2}(\overline{p}_{_{_{\Vert }}}")},  \label{e21}
\end{equation}
where the solution $Z_{_{_{\Vert }}}=0$ has been used.

One can recognize here the same integral that led us to the fermion
condensate in Eq. (\ref{e14}). Therefore, we can write 
\begin{equation}
\mathop{\textstyle \sum }%
\nolimits^{T}\simeq \frac{1}{\pi ^{2}}\frac{\lambda _{y}^{2}}{\lambda
\varphi _{c}^{2}}gBm\ln \left( \frac{gB}{m^{2}}\right) =-2\frac{\lambda
_{y}^{2}}{\lambda \varphi _{c}^{2}}<\overline{\psi }\psi >,  \label{e22}
\end{equation}

Note that the tadpole term is proportional to the magnetic field and
inversely proportional to the scalar mass. We shall see below that such a
functional dependence will be responsible for a visible increment of the
mass solution in this model as compared to other theories.

Taking into account all the leading contributions to Eqs. $\left( \ref{e3}%
\right) $ and $\left( \ref{e4}\right) $ at large magnetic field, one arrives
at the following minimum equations for the fermion mass and the scalar vev
respectively, 
\begin{equation}
m\simeq m_{0}+\left( \frac{g^{2}}{4\pi }-\frac{\lambda _{y}^{2}}{8\pi }%
\right) \frac{m}{4\pi }\ln ^{2}\left( \frac{gB}{m^{2}}\right) +\frac{1}{\pi
^{2}}\frac{\lambda _{y}^{2}}{\lambda \varphi _{c}^{2}}gBm\ln \left( \frac{gB%
}{m^{2}}\right)  \label{e23}
\end{equation}
\begin{equation}
\frac{\lambda }{6}\varphi _{c}^{3}+\frac{\lambda ^{2}}{64\pi ^{2}}\varphi
_{c}^{3}\left( \ln \left( \frac{\varphi _{c}^{2}}{gB}\right) -\frac{11}{3}%
\right) -\lambda _{y}\frac{gB}{2\pi ^{2}}m\ln \left( \frac{gB}{m^{2}}\right)
\simeq 0  \label{e24}
\end{equation}

Eq. $\left( \ref{e24}\right) $ can be simplified by noting that if we assume
that $\varphi _{c}\ll \sqrt{gB},$ one can neglect the term proportional to $%
\lambda ^{2}$ coming from the one-loop scalar self-interaction, compared to
the term coming from the fermion condensate contribution $\sim $ $<\overline{%
\psi }\psi >$. Then, the scalar minimum satisfies

\begin{equation}
\varphi _{c}^{3}\simeq \frac{\lambda _{y}}{\lambda }\frac{3gB}{\pi ^{2}}m\ln
\left( \frac{gB}{m^{2}}\right)  \label{e25}
\end{equation}

Likewise, the contributions of the bubble diagrams to the gap equation,
(second term in $\left( \ref{e23}\right) $), are small compared to the
tadpole contribution, (third term in $\left( \ref{e23}\right) $). Hence,
equation $\left( \ref{e23}\right) $ becomes

\begin{equation}
m\simeq m_{0}+\frac{1}{\pi ^{2}}\frac{\lambda _{y}^{2}}{\lambda \varphi
_{c}^{2}}gBm\ln \left( \frac{gB}{m^{2}}\right)  \label{e26}
\end{equation}

Substituting with Eq. $\left( \ref{e25}\right) $ in $\left( \ref{e26}\right)
,$ we find 
\begin{equation}
m\simeq \frac{1}{\sqrt{\kappa }}\sqrt{gB}  \label{28}
\end{equation}
where the coefficient $\kappa $ satisfies

\begin{equation}
\kappa \ln \kappa \simeq 1.4\frac{\lambda }{\lambda _{y}^{4}}  \label{e29}
\end{equation}

The corresponding solution for the scalar vev is 
\begin{equation}
\varphi _{c}\approx \frac{0.8}{\kappa ^{1/2}\lambda _{y}}\sqrt{gB}
\label{e30}
\end{equation}

Notice that the solutions $\left( \ref{28}\right) $ and $\left( \ref{e30}%
\right) $ are indeed non-perturbative in the couplings constants. At each
fixed scalar self-coupling $\lambda ,$ the values of $m$ and $\varphi _{c}$
increase with $\lambda _{y},$ since the parameter $\kappa $ falls down much
more rapid than $\frac{1}{\lambda _{y}}$ as $\lambda _{y}$ increases.

It is a well known fact that in the absence of a magnetic field, the
one-loop effective action (effective potential) of the present model would
have a minimum at some non-trivial value of the scalar vev, but this minimum
would lie far outside the expected range of validity of the one-loop
approximation, even for arbitrarily small coupling constant, so it would
have to be rejected as an artifact of the used approximation \cite
{coleman-weinb}. As we have just discussed, when a magnetic field is
present, the term $\frac{\partial \Gamma _{f}^{(1)}}{\partial \varphi _{c}},$
Eq.(\ref{e15}), being proportional to the fermionic condensate, dominates
the non-perturbative radiative corrections in Eq.(\ref{e24}). Accordingly, a
non zero scalar field minimum $\left( \ref{e30}\right) ,$ which, as can be
explicitly checked out, is in agreement with the used approximation $\varphi
_{c}\ll \sqrt{gB}$, arises. In other words, thanks to the magnetic field, a
consistent scalar field minimum solution is generated by non-perturbative
radiative corrections. In this sense, a sort of non-perturbative
Coleman-Weinberg mechanism takes place, with the difference that in the
present case no dimensional transmutation occurs. Since the theory already
contains a dimensional parameter: the magnetic field $B$, there is no need
to include scalar-gauge interactions in order to trade a dimensionless
coupling for the dimensional parameter $\varphi _{c}$ \cite{coleman-weinb}.
No constraint between the couplings has to be assumed, except that they are
all sufficiently weak to justify the used Hartree-Fock approximation for the
effective action.

It is clear from the obtained results that the scalar field interactions
significantly enhance the magnetic catalysis. This is better understood
comparing the dynamical mass found in this work with masses generated by
magnetic catalysis in models without scalars. For instance, in QED one has $%
m\simeq \sqrt{gB}\exp (-\frac{\pi }{2}\sqrt{\frac{\pi }{2\alpha }}),$ which
is much smaller mass than $\left( \ref{28}\right) $ (for $\lambda _{y}=0.7$
and $\lambda =0.4,$ which correspond to the top quark Yukawa and Higgs mass $%
M_{H}=115Gev,$ the dynamical mass in the theory with scalar fields is nine
orders of magnitude larger). The enhancement of the dynamical mass due to
the scalar field interactions is comparable to the effect produced by
lowering the number of spatial dimensions in a theory that already exhibits
magnetic catalysis in 3+1 dimensions. For example, in the NJL model the
magnetically catalyzed dynamical mass is visibly increased from $m^{2}=\frac{%
\left| eB\right| }{\pi }\exp \left( -\frac{4\pi ^{2}\varkappa }{\left|
eB\right| }\right) ,$ with $\varkappa =\frac{1}{N_{c}G}-\frac{\Lambda ^{2}}{%
4\pi ^{2}},$ in the (3+1)-dimensional case to $m^{2}=\left| eB\right| ^{2}%
\frac{N_{c}^{2}G^{2}}{\pi }$ in the (2+1)-dimensional one \cite{NJL-2dimen}.

A noteworthy feature of our results is that there is no trivial solution
(stable or unstable) for the scalar vev in the present theory. Besides, no
critical value of the magnetic field is required to produce the fermionic
condensate and the scalar vev.

When studying this model at finite temperature, it is logical to expect that
the increase in the dynamical mass due to the scalar field interactions will
lead to a corresponding increase in the temperature at which the discrete
symmetry is restored. As known, the critical temperature separating chiral
broken-unbroken phases is typically of the order of the zero-temperature
dynamical mass.

We can advance, although it is more or less obvious, that if the current
model is extended to include a complex scalar instead of a real one, none of
the main results will be qualitatively changed. Such a model, however, would
resemble a simplified version of the electromagnetic sector of the $%
SU(2)\times U(1)$ electroweak model (the scalar would not be coupled to the
electromagnetic field, just as the Higgs field of the electroweak theory).
Since no tree-level scalar mass term with the wrong sign would be introduced
in such a theory, the external magnetic field would catalyze gauge symmetry
breaking by producing a non-trivial scalar vev through the non-perturbative
mechanism reported in this paper.

It is still too premature to envision all possible practical implications of
the present work. It seems to us however that a careful investigation in the
context of the electroweak theory of the non-perturbative effects here
discussed is required. In the high temperature region of the electroweak
theory, where the temperature-dependent fermion masses are nearly zero,
these magnetic field-driven non-perturbative effects may be important at
reasonable large magnetic fields. An interesting point to consider would be
whether the new magnetic field dependence of the scalar vev would change the
recent conclusions on the role of magnetic fields in electroweak
baryogenesis \cite{Giovannini-Shaposhnikov}.

\begin{acknowledgments}
It is a pleasure to thank V. Gusynin and V. Miransky for enlightening
discussions on the phenomenon of magnetic catalysis, and M. Laine, M.
Shaposhnikov and G. Veneziano for useful discussions and remarks. This work has been
supported in part by NSF grant PHY-9722059 (EF and VI) and NSF POWRE grant
PHY-9973708 (VI).
\end{acknowledgments}

\end{document}